\documentclass{aa}  
\usepackage{graphicx}
\usepackage{txfonts}
\usepackage{natbib}

\begin{document}

\topmargin=-32.0mm

\title{On posterior probability and significance level: application to the power spectrum of HD49933 observed by CoRoT\thanks{The CoRoT space mission, launched on
   2006 December 27, was developed and is operated by the CNES, with
   participation of the Science Programs of ESA, ESA's RSSD, Austria,
   Belgium, Brazil, Germany and Spain.}}
   
   \titlerunning{On posterior probability and significance level}

\author{T.~Appourchaux\inst{1}, R.~Samadi\inst{2}, M.-A.~Dupret\inst{2,3}}

\institute{Institut d'Astrophysique Spatiale, UMR8617, Universit\'e Paris XI, B\^atiment 121, 91405 Orsay Cedex, France\\
              \email{Thierry.Appourchaux@ias.u-psud.fr}
              \and
            Observatoire de Paris, LESIA, UMR8109, 92195 Meudon Cedex, France
            \and
            Institut d'Astrophysique et de G\'eophysique de l'Universit\'e de Li\`ege, All\'ee du 6 Ao\^ut 17 Ð B 4000 Li\`ege, Belgium
}

\abstract
{}
{We emphasize that the mention of the significance level when rejecting the null hypothesis (H$_{0}$ which assumes that what is observed is pure noise) can mislead one to think that the H$_{0}$ hypothesis is unlikely to occur with that significance level.  We show that the significance level has nothing to do with the posterior probability of H$_{0}$ given the observed data set, and that this posterior probability is very much higher than what the significance level naively provides.}
{We use Bayes theorem for deriving the posterior probability of H$_{0}$ being true assuming an alternative hypothesis H$_{1}$ that assumes that a mode is present, taking some prior for the mode height, for the mode amplitude and linewidth.}
{We report the posterior probability of H$_{0}$ for the p modes detected on HD49933 by CoRoT.}
{We conclude that the posterior probability of H$_{0}$ provide a much more conservative quantification of the mode detection than the significance level.  This framework can be applied to any stellar power spectra similar to those obtained for asteroseismology.}


   \maketitle

\section{Introduction}
In the field of helioseismology, the null hypothesis H$_{0}$ has been used by \citet{TA2000} to infer upper limits on the amplitude of g modes.  The H$_{0}$ hypothesis assumes that what is observed is pure noise.  This hypothesis has been used for classical variable stars for detecting peaks in a power spectrum \citep{Scargle82}. In these papers, the authors accepted the H$_{0}$ and set an upper limit corresponding to a threshold level of say 10\%.  Here we argue, that this threshold has been somewhat arbitrarily chosen {\it a priori}.  By accepting or rejecting the H$_{0}$ hypothesis, there is no discussion as to why borderline case should be rejected or accepted.  This abrupt truncation between the {\it good} and the {\it bad} leads to a decision that could have been different if the threshold had been different.

In the following sections, we first lay down the foundation for understanding the meaning of the H$_0$ hypothesis, and then explain what is commonly misunderstood about the H$_{0}$ hypothesis.  We then derive, for specific cases encountered in helio- and asteroseismology, the posterior probability of H$_{0}$.  We show how one can use the formalism for real data such as that gathered by the CoRoT mission for HD49933 and then conclude.

\section{Significance level and the H$_{0}$ hypothesis} 
\citet{Fisher} devised the well known Fisher test for testing the null hypothesis (H$_0$).  For that test, a threshold of 5\% is commonly used; the p-value quoted is the value of the test if it is less than this threshold.  So, for instance, a result of 4.9\% would result in rejecting the H$_{0}$ hypothesis, while 5.1\% would result in accepting the H$_{0}$ hypothesis.  When the H$_{0}$ hypothesis is rejected, the reported p-value is used as a significance for the validity of not accepting the H$_{0}$ hypothesis.  The so-called borderline cases led in the medical field to findings related to effectiveness of medicine that were, sometimes, not proven by subsequent studies.  The controversy about the use of p-values that occurred in the medical world is directly related to the abrupt and arbitrary cut-off of  the threshold applied (be 5\% or 10\%).  Although the relevance of what has been found in the medical field could seem remote to most astrophysicists, it is indeed extremely relevant to understand that the improper use of p-values is the same as the improper use of the so-called {\it significance level}.

Reporting a small number for the significance level should not be used for claiming the proper rejection of the H$_{0}$ hypothesis.  The mistake is to ascribe a significance level to a measurement carried out only once, not repeated and spanning just a very small volume of the space of the parameters.  When making an observation of random variable $x$, one wants to check the probability that what is observed could be due to noise.  For that purpose, a test statistic is derived $T(x)$.  If one observes a value of $x=X$ that would not reject the H$_{0}$, then one compute, $p$, the significance level or p-value, defined as:
\begin{equation}
	p = P_0(T(x) \geq T(X))
\end{equation}
where $P_0$ is the probability of having $T(x) \geq T(X)$ when H$_0$ is true.  The test set by Eq. (1) is about checking that the statistical test $T(x)$ is compared with the value given by $T(X)$.  When computing the statistical test $T(x)$, one does not span the space of $X$: one has $x=X$: this is the so-called point null hypothesis.  In other terms, what are the probabilities that one has {\it exactly} that value of $x$?  This is a completely different question from knowing how true H$_{0}$ is, i.e. obtaining $p({\rm H}_{0} | x)$.   \citet{Berger1987} provided a way of deriving $p({\rm H}_{0} | x)$ with respect to $p(x | {\rm H}_0)$ and $p(x | {\rm H}_1)$, where H$_{1}$ is the alternative hypothesis (i.e. there is a signal).  {\bf They obtained using Bayes theorem, the so called posterior probability of H$_{0}$ given the observed data $x$:
\begin{equation}
p({\rm H}_{0} | x) =\frac{ p({\rm H}_{0}) p( x | {\rm H}_{0})}{p({\rm H}_{0}) p( x | {\rm H}_{0})+p({\rm H}_{1}) p( x | {\rm H}_{1})}
\end{equation}
where $p({\rm H}_{0})$ and $p({\rm H}_{1})$ are the probabilities ascribed to the H$_0$ and H$_1$ hypothesis, respectively.  Note that what we really want to know is $p({\rm H}_{0} | x)$, the probability of H$_0$ being true given the data we have, not the probability of the data $x$ given H$_{0}$, i.e. $p(x | {\rm H}_{0})$.  In order to minimize the impact of the probabilities assigned to both hypothesis, we assume that these are equally probable ($p_0$=0.5)}:
\begin{equation}
p({\rm H}_{0} | x)= \left(1+\frac{p(x | {\rm H}_1)}{p(x | {\rm H}_0)}\right)^{-1}
\end{equation}
{\bf Under these assumptions} \citet{Sellke2001} found that the probability $p({\rm H}_{0} | x)$ of having H$_{0}$ being true given some observed data $x$ of a random variable $X$ has a lower bound:
\begin{equation}
  p({\rm H}_{0} | x) \geq \left(1-\frac{1}{{\rm e} p \ln p}\right)^{-1}
\end{equation}
An immediate consequence is that for a significance level of 1\%, the odds against H$_0$ are at least 10 to 1, and for 10\%, the odds against H$_0$ are at least 2.6 to 1.  In both case, the likelihood of wrongly rejecting H$_0$ is much higher than what the p-value provides, by at most a factor 10 and 4 respectively.  

\citet{Sellke2001} were able to set a lower value to $p({\rm H}_{0} | x)$ for almost an arbitrary alternative H$_{1}$ hypothesis.  We show that there is  indeed a lower bound when one wants to detect peaks in power spectra.  Hereafter, we give several examples of how one can derive in practice the posterior probability of H$_{0}$.

\section{Posterior probability for peak detection}
\subsection{Long lived modes}
\subsubsection{Mode height known a priori}
Here we take the case of a power spectrum for which we seek a peak restricted to a single frequency bin.  The power spectrum has a $\chi^2$ with 2 d.o.f. statistics, for which a bin has reached a value $x$.  We want to test if this could be the result of a true sine wave or due to noise.  We have for the H$_{0}$ hypothesis:
\begin{equation}
p(x | {\rm H}_0)={\rm e}^{-x}
\end{equation}
The noise is assumed to be 1.  For the alternative hypothesis H$_{1}$, we assume that there is a signal of a long lived mode, i.e. restricted to one bin, for which the mode height $H$ is known and the mode is stochastically excited (like a stellar p mode).  Then we have:
\begin{equation}
p(x | {\rm H}_1)=\frac{1}{1+H}{\rm e}^{-x/(1+H)}
\end{equation}
{\bf Equation (3) is then rewritten for our problem as}:
\begin{equation}
p({\rm H}_{0} | x)= \left(1+\frac{1}{1+H}{\rm e}^{xH/(1+H)}\right)^{-1}
\end{equation}
Since the significance $p=p(x | {\rm H}_0)$, we finally have:
 \begin{equation}
p({\rm H}_{0} | x)= \left(1+\frac{1}{1+H}p^{-H/(1+H)}\right)^{-1}
\end{equation}
It can be shown that the minimum of $p({\rm H}_{0} | x)$ is reached for $H=-\ln p-1$, and with the value:
 \begin{equation}
p_{\rm min}^{H}({\rm H}_{0} | x)= \left(1-\frac{1}{{\rm e} p \ln p}\right)^{-1}
\end{equation}
In that case the lower bound set by {\bf Eq. (4)} is reached.

\subsubsection{Mode height unknown}
Of course, most of the time one does not {\it know} the height of the mode to be detected.  We can assume a prior for the mode height (uniform distribution, gaussian, etc...).  For example if we assume that the mode height is uniformly distributed over some range {\bf [0,$H_{\rm u}$]}, $p(x  | {\rm H}_1)$ is rewritten as:
\begin{equation}
p^{\rm uni}(x | {\rm H}_1)=\frac{1}{H_{\rm u}}\int_0^{H_{\rm u}} \frac{1}{1+H'}{\rm e}^{-x/(1+H')} {\rm d}H'  
\end{equation}
Then {\bf Eq. (3)} can be rewritten as:
\begin{equation}
p^{\rm uni}({\rm H}_{0} | x)= \left(1+\frac{1}{H_{\rm u}}\int_0^{H_{\rm u}} \frac{1}{1+H'}p^{-H'/(1+H')} {\rm d}H'\right)^{-1}
\end{equation}
It can be shown that there is a minimum reached at a value of $H_{\rm min}$ by solving:
\begin{equation}
\frac{1}{H_{\rm min}}\int_0^{H_{\rm min}} \frac{1}{1+H'}p^{-H'/(1+H')} {\rm d}H' = \frac{1}{1+H_{\rm min}}p^{-H_{\rm min}/(1+H_{\rm min})}
\end{equation}
The minimum is then given by:
\begin{equation}
p_{\rm min}^{\rm uni}({\rm H}_{0} | x)= \left(1+\frac{1}{1+H_{\rm min}}p^{-H_{\rm min}/(1+H_{\rm min})}\right)^{-1} 
\end{equation}
From Eqs. (8) and (9), we can deduce that
\begin{equation}
p_{\rm min}^{\rm uni}({\rm H}_{0} | x) >  \left(1-\frac{1}{{\rm e} p \ln p}\right)^{-1}
\end{equation}
It means that when there is more uncertainty about the possible height of the mode, one is less likely to reject the ${\rm H}_{0}$ hypothesis.

\begin{figure}[!]
\centering
{
\includegraphics[width=6 cm,angle=90]{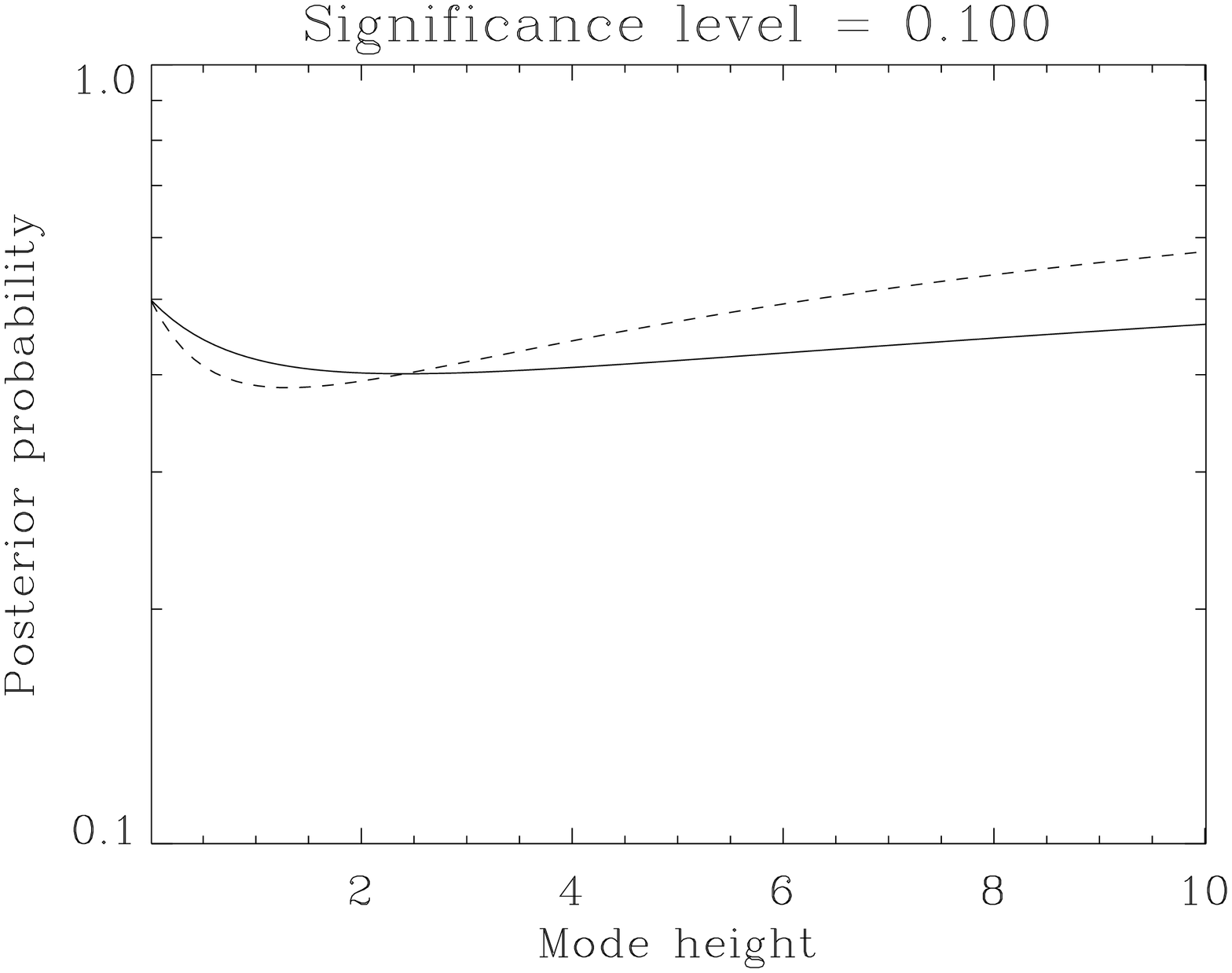}
\includegraphics[width=6 cm,angle=90]{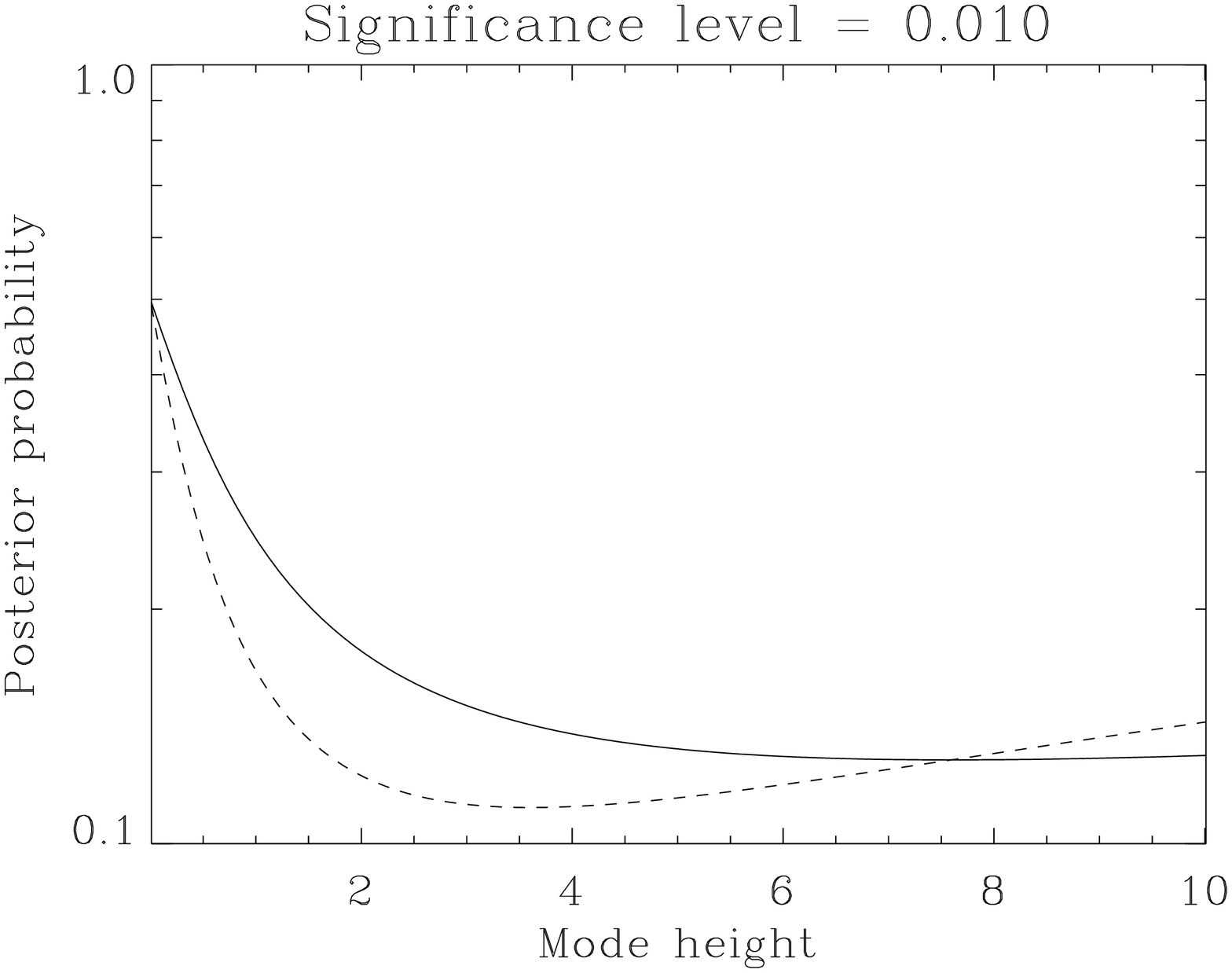}
}   
\caption{Posterior probability as a function of the known mode height (dashed line), or as a function of {\bf the mode height range ($H_{\rm u}$)} of the uniform prior (continuous line) for a significance level of 10\%.  For the known mode height, the minimum is reached for an height of 1.3 with a value of 38\%.  For the unknown mode height, the minimum is reached at a value of 40\%.  Posterior probability as a function of the known height (dashed line), or as a function of the maximum mode height of the uniform prior (continuous line) for a significance level of 1\%.  For the known mode height, the minimum is reached for an height of 3.6 with a value of 11\%.  For the unknown mode height, the minimum is reached at a value of 12.7\%.}
\label{}
\end{figure}

\begin{figure}[!]
\centering
{
\includegraphics[width=6 cm,angle=90]{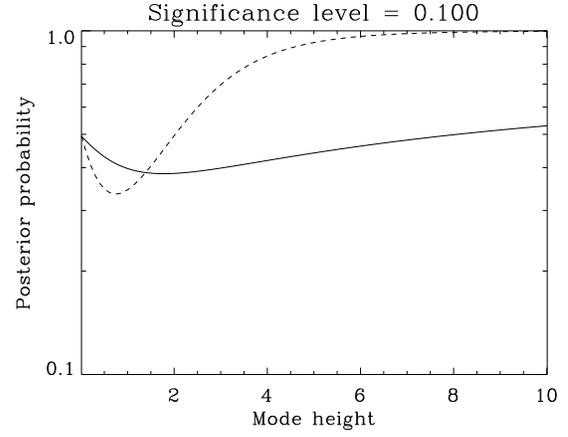}
}   
\caption{Posterior probability for a short lived mode when smoothing the power spectra over 10 bins corresponding to a window twice as large as the mode linewidth.  The posterior probability is given as a function of the known mode height (dashed line), or as a function of the {\bf mode height range ($H_{\rm u}$)} of the uniform prior (continuous line) and a uniform prior for the linewidth, for a significance level of 10\%.}
\label{}
\end{figure}

\subsection{Short lived modes}
 \citet{Appourchaux2004} reported on how one can detect a mode having a lifetime shorter than the observation time. .  He suggested smoothing the power spectrum in order to enhance the signal-to-noise ratio and provided the analytical expression for H$_{0}$ related to the summation over $n$ bins of a $\chi^2$ statistic with constant mean; and for H$_{1}$ related to the summation over $n$ bins of a $\chi^2$ statistic having different means due to the presence of the mode profile.  Therefore, when the spectrum is summed over $n$ bins, we can derive for the H$_{0}$ hypothesis :
\begin{equation}
p(x | {\rm H}_0)=\frac{x^{n-1}{\rm e}^{-x}}{\gamma(n)} 
\end{equation}
where the mean of the power spectrum is 1, while $n$ is the mean of the smoothed power spectrum (for simplicity, we assumed that $S=1$), and $\gamma(n)$ is the Gamma function. The significance level {\bf $x$} is then given by {\bf solving}:
\begin{equation}
	p=\frac{1}{\gamma(n)}\int_{x}^{+\infty} u^{n-1} {\rm e}^{-u}{\rm d}u
	\label{first}
\end{equation}
Assuming, that the mode has a known amplitude $A$ and a known linewidth $\Gamma$, we can write using Eq. (8) of \citet{Appourchaux2004} the following approximation:
\begin{equation}
p(x | {\rm H}_1)=\frac{\lambda^{\nu}}{\gamma(\nu)} {x}^{\nu-1}{\rm e}^{-\lambda {x}}
\end{equation}
where $\lambda$ and $\nu$, {\bf given in \citet{Appourchaux2004} are obtained by integrating symmetrically around the central frequency of the mode, thereby ensuring that the signal is maximum; $\lambda$ and $\nu$ are functions of the mode height $H (=A^2/\pi \Gamma)$ and $\Gamma$}.  If we do not know the mode amplitude and its linewidth, we can specify a prior for the amplitude and the linewidth {\bf which, for instance, can be done using a model of mode excitation}.  Here we assume that the mode amplitude $A$ and the linewidth $\Gamma$ are independent (this would not be the case of $H$ and $\Gamma$).  If we use uniform priors, we then have:
\begin{equation}
p(x | {\rm H}_1)=\frac{1}{A_{\rm u} \Gamma_{\rm u}} \int_0^{A_{\rm u}} \int_0^{\Gamma_{\rm u}}\frac{\lambda^{\nu}}{\gamma(\nu)} {x}^{\nu-1}{\rm e}^{-\lambda {x}} {\rm d}A'{\rm d}\Gamma'
\end{equation}
{\bf This equation would be similar if we were to have a uniform prior on the mode height, $A$ would then be replaced by $H$.}  Combining Eqs. (16) and (17) in {\bf Eq. (3)}, we can then obtain the posterior probability $p( {\rm H}_0 | x)$ when the mode amplitude and linewidth are known.  If we replace Eq. (17) by Eq. (18), we then obtain the probability $p( {\rm H}_0 | x)$ when the mode amplitude and linewidth are not known.

\subsection{Discussion}
Figures 1 shows the posterior probability for long lived modes for a known mode height (Eq. 8) and an unknown mode height (Eq. 11) for two different significance levels.  {\bf Figure 2} shows the posterior probability for short lived modes for a known mode height and linewidth {\bf (Eqs. 15 and 17)} and for an unknown mode height and linewidth {\bf (Eqs. 15 and 18, {\bf with $H$ replacing $A$})} for a single significance level; even a significance level of 1\% does not provide a better rejection of the H$_0$ hypothesis.   These posterior probabilities have a lower bound which means that even a very low significance level is no guarantee for positive detection!  

It is also counter intuitive that the posterior probability increases when the mode height (known or unknown) increases.  We recall that the significance level $p$ corresponds to the level $x$ at which the peak has been observed (i.e. it is $p=e^{-x}$ for long lived modes).  If we assume a priori that the mode has a high mode height, then the observation at a low significance level indicates that our assumption on the high mode height is not correct, and that the data dismisses the a priori made on the mode height.  In other words, it is more probable that the H$_0$ hypothesis is true.

In case of the absence of an alternative hypothesis, it is better to set a low significance value that will result in a likely rejection of the H$_0$ hypothesis.  Figure 3 shows the lower bound\footnote{The same lower bound as Eq. (9) for a known mode height} set by Eq. (4) compared to the minimum found {\bf using Eq. (11)} for the uniform prior on the mode height .


\begin{figure}[!]
\centering
{\includegraphics[width=6 cm,angle=90] {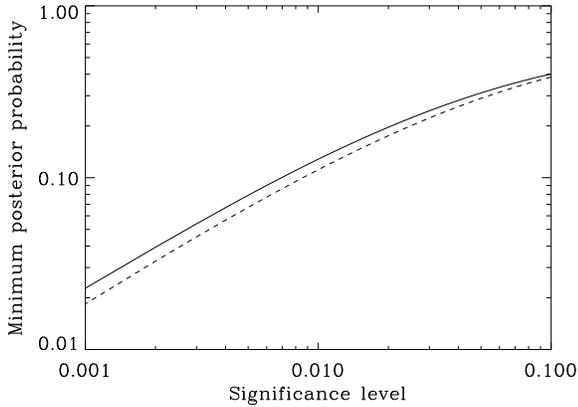}}
\caption{Lower bound to the posterior probability as function of the significance level for the known mode height (dashed line) and for the uniform prior (solid line).}
\label{}
\end{figure}

\section{Application to the CoRoT data: HD49933}
The computation of the posterior probability has been applied to CoRoT for illustrative purposes.  The data used are those of the first initial run performed on HD49933 \citep{Appourchaux2008}.  The objective was to provide an objective way of detecting oscillation modes in HD49933 that could be applied to any other star.

The methodology used for deriving the posterior probability is as follows:
\begin{itemize}
\item we compute the power spectrum from the detrended time series, as in \citet{Appourchaux2008}.
\item we smooth the spectrum over $n$ bins using a boxcar.
\item we select 30 50-$\mu$Hz wide windows starting at 1200 $\mu$Hz (the 50-$\mu$Hz window corresponds
	roughly to half the large frequency separation).
\item For each window, we compute the median in the window of the smoothed spectrum 
         (it provides an estimate of the mean noise level were the modes not present).
\item The smoothed spectrum is normalized in each window by dividing by the median and multiplying by the
	number of smoothing bins $n$ (it provides values commensurate with these of Eq. 15).
\item we apply the H$_0$ hypothesis using a detection probability of a signal being due to noise of 10\% over all the 30 windows, taking into account the fact that in each window the number of independent bins is 50 ${\delta \nu}^{-1} n^{-1}$ ($\delta \nu$ is the frequency resolution of the original power spectrum).  The detection probability is then 0.1 $(N_w)^{-1} ( 50 {\delta \nu}^{-1} n^{-1})^{-1}$ per independent bin.
\item we then solve Eq. (16) for $x_{\rm det}$ given the detection level given above.
\item In each window, we then select  the bins that are greater than $x_{\rm det}$, i.e. we accept or reject the H$_{0}$ hypothesis.
\item After the selection, we keep the greatest value $x_{\rm max}$ found in the window {\bf corresponding to the central frequency of the mode (See Eq. 17).}
\item we then compute the posterior probability of H$_{0}$ given by {\bf Eq (3)} using Eqs. (15) and (18) assuming some prior on the mode height and linewidth as described below.  
\item For comparison, we also compute the significance level as given by Eq. (16) from the value of $x_{\rm max}$.
\end{itemize}

The theoretical amplitudes for HD49933 are derived from \citet{Samadi2009a} using an adiabatic treatment of radiative transfer, and the excitation rate as calculated in \citet{Samadi2009b}.  The theoretical linewidths have been computed with the non-adiabatic pulsation code MAD. This code includes a time-dependent convection treatment described in \citet{Griga2005}: it takes into account the role played by the variations of the convective flux, the turbulent pressure, and the dissipation rate of turbulent kinetic energy.  This treatment is non-local, with three free parameters $a$, $b$, and $c$ corresponding to the non-locality of the convective flux, the turbulent pressure and the entropy gradient.  We take here the values $a=10$, $b=3$, and $c=3.5$ obtained by fitting the convective flux and turbulent pressure of 3D hydrodynamic simulations in the upper overshooting region of the Sun \citep{MAD06a}.  According to \cite{Griga2005}, we introduced a free complex parameter $\beta$ in the perturbation of the energy closure equation.  We use here the value $\beta=-3i$, which leads to a good agreement between the theoretical and observed linewidths and phase lags in the range of solar pressure modes. A 1D stellar model obtained with the code CESAM was used for our non-adiabatic computations.  It has a solar metallicity and reproduces the effective temperature and gravity of HD 49933 \citep{Samadi2009a}.  The amplitudes and linewidths are shown on Figure 4.  We have linearly interpolated both curves to provide a continuous prior with frequency.  

The uniform priors for amplitude and linewidth are derived from the theory by taking into account of an {\it uncertainty factor} in the theoretical model.  For amplitude, we assumed that the maximum is $\sqrt{2}$ larger than given on Fig. 4 (twice in energy); for linewidth, we assumed that the maximum is twice  greater than given on Fig. 4.  Note that a larger prior increases the posterior probability, as shown on Figs. 1 and 2.  We assumed that the noise floor in HD49933 is given by the photon noise, which is about 0.15 ppm$^{2}/\mu$Hz \citep{Appourchaux2008}.

Figure 5 shows the results of the procedure described above.  It is clear that the posterior probability is higher than the significance level, i.e. the posterior probability provides a more conservative number (H$_{0}$ more likely).  The smoothing procedure also shows two effects that were predicted by \citet{Appourchaux2004}: first, short lived modes are easier to detect when the spectrum is smoothed, second, long lived modes are more difficult to detect when the spectrum is smoothed.  The first effect manifests itself in the larger number of detected modes at higher frequency and by the decrease of the prior probability (i.e. the signal is more likely).  The second effect is seen at low frequency where a couple of modes have their prior probability increased to non-negligible value after smoothing (i.e. the signal is less likely).  When we compare with modes reported by \citet{Appourchaux2008}, we find that more than 85\% of the $l=0-2$ mode pairs and $l=1$ modes are recovered.  An additional mode at 2579 $\mu$Hz is detected that could be an $l=1$ according to the identification of \citet{Appourchaux2008}.

\begin{figure}[!]
\centering
{
\includegraphics[width=6.5 cm,angle=90]{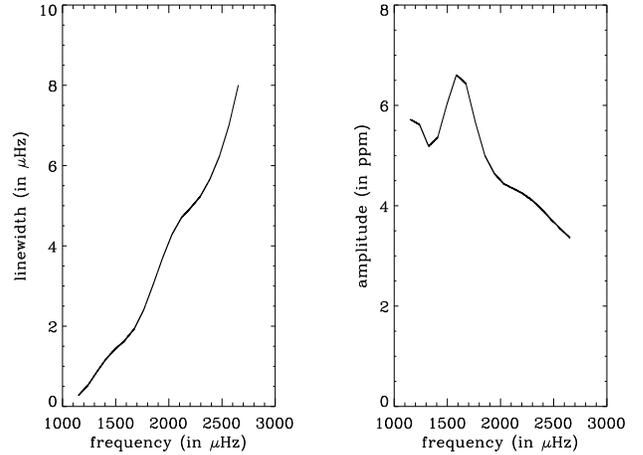}
}   
\caption{(Left) Theoretical mode linewidth of HD49933 as a function of frequency.  (Right) Theoretical mode amplitude of HD49933 as a function of frequency.}
\label{}
\end{figure}


\begin{figure}[!]
\centering
{
\includegraphics[width=6 cm,angle=90]{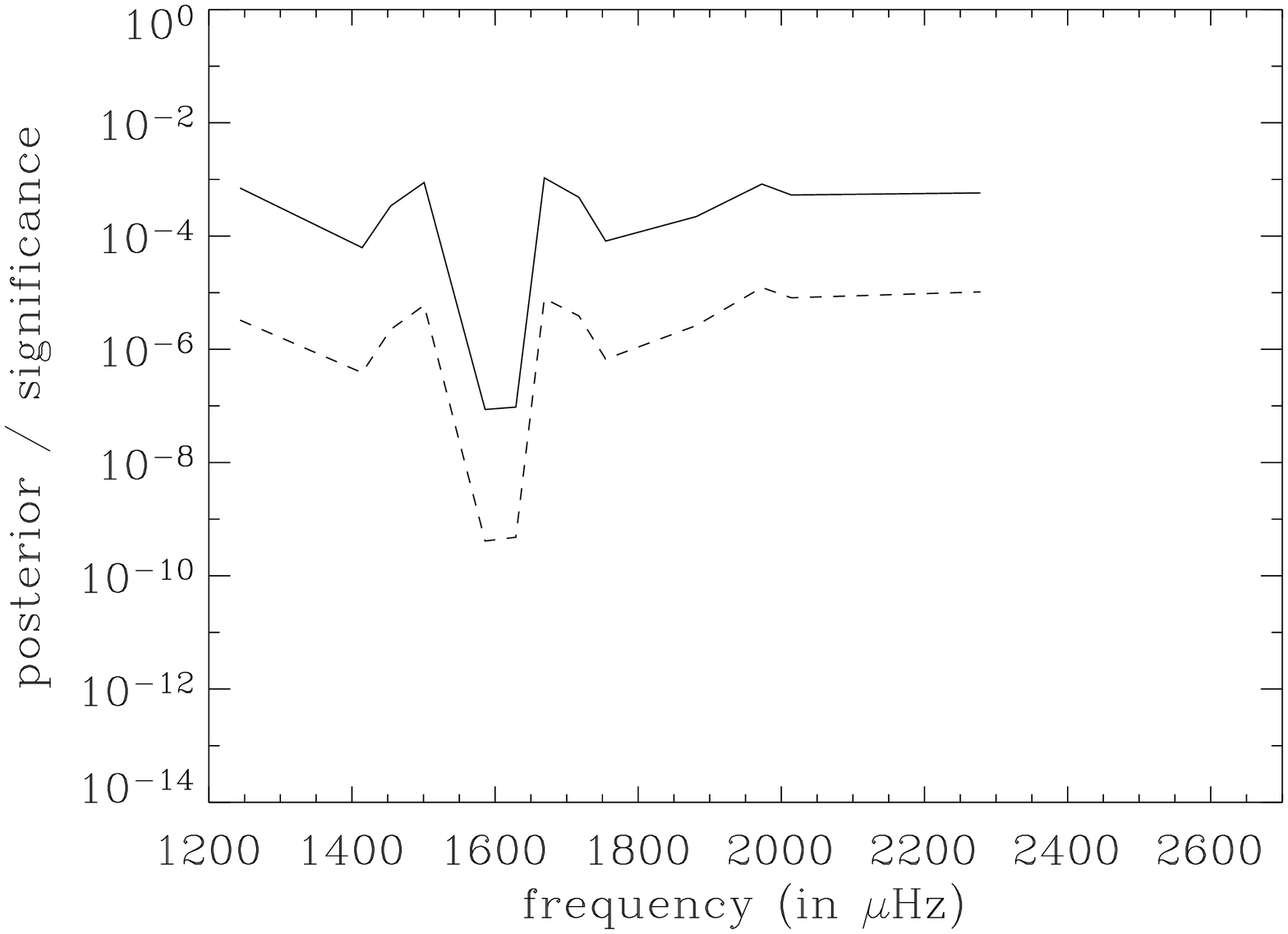}
\includegraphics[width=6 cm,angle=90]{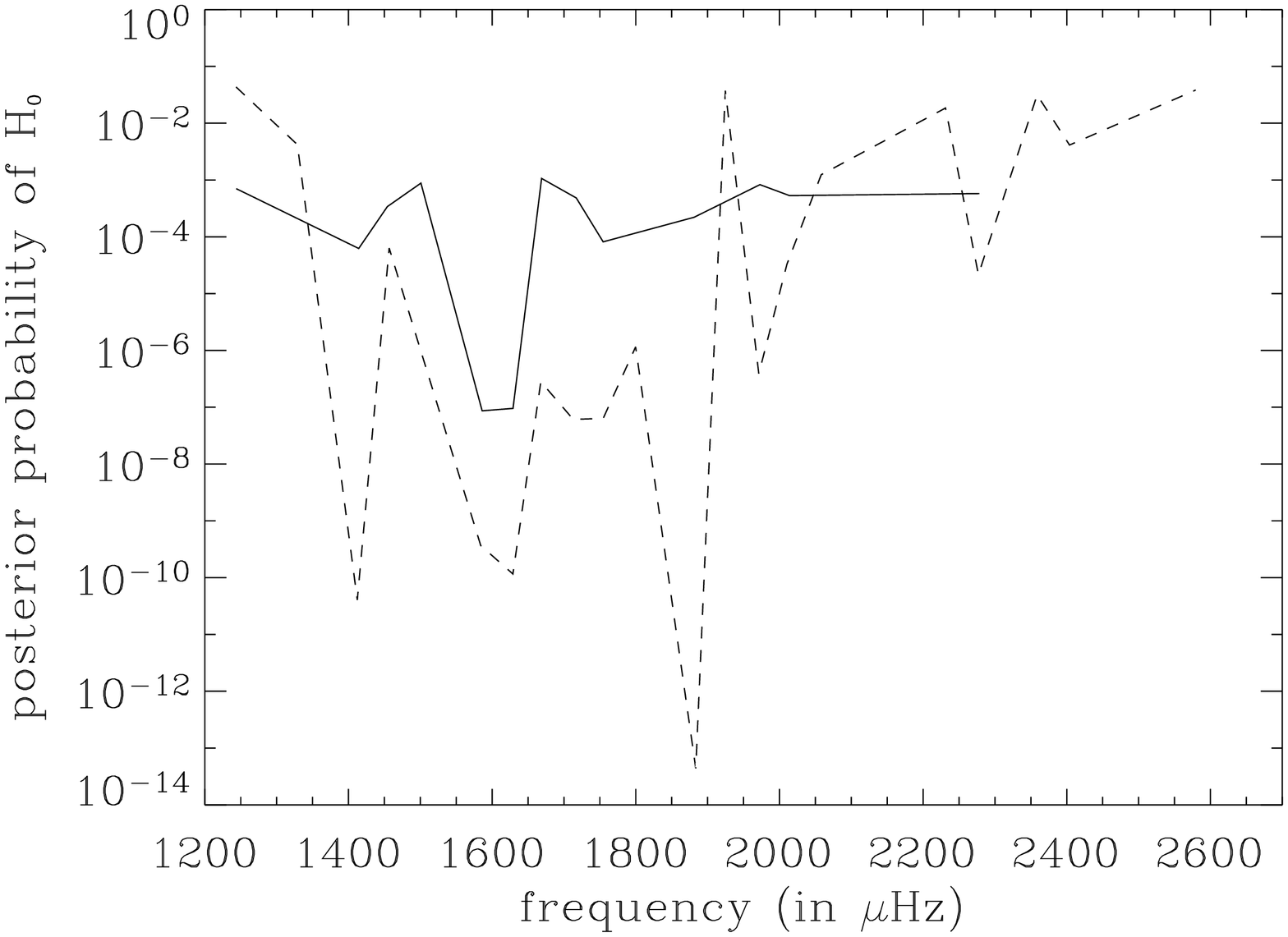}
}   
\caption{(Top) Comparison of the posterior probability (solid line) to the significance level (dashed line) as a function of frequency for a spectrum smoothed over 5 bins ($\approx 1 \mu$Hz). (Bottom) Posterior probability as a function of frequency for a spectrum smoothed over 5 bins (solid line) and over 50 bins (dashed line), respectively $\approx 1 \mu$Hz and $\approx 10 \mu$Hz}
\label{}
\end{figure}

\section{Conclusion}
The significance level refers to the significance of the data {\it given} the hypothesis, while we are interested in the posterior probability of the null hypothesis {\it given} the data.  Here we showed that for a significance level of 10\%, the posterior probability of the null hypothesis is at least 38\% when there is no alternative hypothesis.  We have shown how one can in practice calculate and compute the posterior probability for the null hypothesis.  This has been applied to several examples and to the CoRoT data.  We have shown, for the first time, how one can assess the detectability of short lived p modes in a power spectrum.  The methodology can be applied to any stellar power spectrum for which theoretical expectations are available.

\acknowledgements
We are grateful to John Leibacher for checking the English and for correcting several typos.  {\bf We are thankful to the referee for making the paper clearer.}

\bibliographystyle{aa}
\bibliography{thierrya}

\begin{thebibliography}{11}
\expandafter\ifx\csname natexlab\endcsname\relax\def\natexlab#1{#1}\fi

\bibitem[{{Appourchaux}(2004)}]{Appourchaux2004}
{Appourchaux}, T. 2004, \aap, {\bf 428}, 1039

\bibitem[{{Appourchaux} {et~al.}(2000){Appourchaux}, {Fr\"ohlich}, {Andersen},
  {Berthomieu}, {Chaplin}, {Elsworth}, {Finsterle}, {Gough}, {Hoeksema},
  {Isaak}, {Kosovichev}, {Provost}, {Scherrer}, {Sekii}, \& {Toutain}}]{TA2000}
{Appourchaux}, T., {Fr\"ohlich}, C., {Andersen}, B., {et~al.} 2000, \apj, {\bf
  538}, 401

\bibitem[{{Appourchaux} {et~al.}(2008){Appourchaux}, {Michel}, {Auvergne},
  {Baglin}, {Toutain}, {Baudin}, {Benomar}, {Chaplin}, {Deheuvels}, {Samadi},
  {Verner}, {Boumier}, {Garc{\'{\i}}a}, {Mosser}, {Hulot}, {Ballot}, {Barban},
  {Elsworth}, {Jim{\'e}nez-Reyes}, {Kjeldsen}, {R{\'e}gulo}, \&
  {Roxburgh}}]{Appourchaux2008}
{Appourchaux}, T., {Michel}, E., {Auvergne}, M., {et~al.} 2008, \aap, {\bf
  488}, 705

\bibitem[{{Berger} \& {Sellke}(1987)}]{Berger1987}
{Berger}, J. \& {Sellke}, T. 1987, Journal of the American Statistical
  Association, {\bf 82(397)}, 112

\bibitem[{{Dupret} {et~al.}(2006){Dupret}, {Samadi}, {Grigahcene},
  F1999{Goupil}, \& {Gabriel}}]{MAD06a}
{Dupret}, M.-A., {Samadi}, R., {Grigahcene}, A., F1999{Goupil}, M.-J., \&
  {Gabriel}, M. 2006, Communications in Asteroseismology, 147, 85

\bibitem[{{Fisher}(1925)}]{Fisher}
{Fisher}, R.~A. 1925, {Statistical Methods for Research Workers} (Oliver and
  Boyd, Edinburgh, Scotland), 299

\bibitem[{{Grigahc{\`e}ne} {et~al.}(2005){Grigahc{\`e}ne}, {Dupret}, {Gabriel},
  {Garrido}, \& {Scuflaire}}]{Griga2005}
{Grigahc{\`e}ne}, A., {Dupret}, M.-A., {Gabriel}, M., {Garrido}, R., \&
  {Scuflaire}, R. 2005, \aap, {\bf 434}, 1055

\bibitem[{{Samadi} {et~al.}(2009{\natexlab{a}}){Samadi}, {Ludwig}, {Belkacem},
  {Goupil}, , {Benomar}, {Dupret}, {Baudin}, \& {Appourchaux}}]{Samadi2009a}
{Samadi}, R., {Ludwig}, H.-G., {Belkacem}, K., {et~al.} 2009{\natexlab{a}},
  \aap, to be submitted

\bibitem[{{Samadi} {et~al.}(2009{\natexlab{b}}){Samadi}, {Ludwig}, {Belkacem},
  {Goupil}, \& {Dupret}}]{Samadi2009b}
{Samadi}, R., {Ludwig}, H.-G., {Belkacem}, K., {Goupil}, M., \& {Dupret}, M.-A.
  2009{\natexlab{b}}, \aap, to be submitted

\bibitem[{{Scargle}(1982)}]{Scargle82}
{Scargle}, J.~D. 1982, \apj, 263, 835

\bibitem[{{Sellke} {et~al.}(2001){Sellke}, {Bayarri}, \& {Berger}}]{Sellke2001}
{Sellke}, T., {Bayarri}, M.~J., \& {Berger}, J. 2001, The American
  Statistician, {\bf 55}, 62

\end{thebibliography}

\end{document}